\documentclass[10pt,prl,twocolumn,nofootinbib,superscriptaddress,floatfix]{revtex4-2}
\pdfoutput=1
\usepackage{amsmath, amssymb}
\usepackage{mathtools}
\usepackage{graphicx}
\usepackage{xcolor}
\usepackage[colorlinks=true,linkcolor=blue,citecolor=magenta,urlcolor=magenta]{hyperref}

\newcommand{\Tr}{\operatorname{Tr}}
\newcommand{\DD}{\mathcal{D}_d}
\newcommand{\MM}{\mathcal{M}_d}
\newcommand{\PP}{\mathcal{P}_d}
\newcommand{\Km}{K_{\mu}}
\newcommand{\eps}{\varepsilon}
\newcommand{\stat}{\mathsf{s}}
\newcommand{\Hil}{\mathcal{H}}
\newcommand{\Fock}{\mathfrak{F}}
\newcommand{\dG}{d\Gamma}
\newcommand{\Nhat}{\hat{N}}
\newcommand{\Cov}{\operatorname{Cov}}
\newcommand{\appref}[1]{\hyperref[app:#1]{Appendix~#1}}

\newtheorem{theorem}{Theorem}

\newtheorem{lemma}{Lemma}

\newtheorem{proposition}[theorem]{Proposition}
\newenvironment{proof}{\par\noindent\textit{Proof.}\ }{\hfill$\square$\par\medskip}

\begin{document}

\title{Semidefinite optimization as many-body thermodynamics:\\Boltzmann, Fermi--Dirac, and Bose--Einstein frameworks}

\author{Michele Minervini}
\affiliation{School of Electrical and Computer Engineering, Cornell University, Ithaca, New York 14850, USA}

\author{Nana Liu}
\affiliation{Institute of Natural Sciences, School of Mathematical Sciences, Ministry of Education Key Laboratory in Scientific and Engineering Computing, and Global College, Shanghai Jiao Tong University, Shanghai 200240, China}

\author{Dhrumil Patel}
\affiliation{School of Computer Science, Cornell University, Ithaca, New York 14850, USA}

\author{Mark M. Wilde}
\affiliation{School of Electrical and Computer Engineering, Cornell University, Ithaca, New York 14850, USA}

\begin{abstract}
We study the semidefinite programs (SDPs) of quantum information whose variables are density operators, measurement operators, or unbounded positive semidefinite operators, together with their thermodynamic regularizations by the von Neumann, Fermi--Dirac, and Bose--Einstein entropies, which have been associated with Boltzmann, Fermi--Dirac, and Bose--Einstein statistics, respectively. We find that this correspondence is exact: the Fermi--Dirac and Bose--Einstein regularized SDPs coincide with the constrained free-energy minimization of an ideal Fermi or Bose gas whose Hamiltonian and non-commuting charges are the second quantizations of the SDP data and whose one-body density matrix is the SDP variable, while the Boltzmann framework corresponds to the single-particle case. We show that the three frameworks can be unified under one framework, in which each entropy-regularized dual is, up to a term linear in the chemical potentials, the grand potential of the corresponding system; the duals are unconstrained and concave in the chemical-potential vector, with gradient and Hessian given by thermal expectation values that hybrid quantum--classical algorithms can estimate. We further identify the Fermi--Dirac and Bose--Einstein relative entropies with the Umegaki relative entropy between quasi-free states, which yields their monotonicity under gauge-covariant Gaussian channels. The result places a wide class of SDPs and their solvers within the thermodynamics of ideal quantum gases and points to extensions to interacting gases.
\end{abstract}

\maketitle


\textit{Introduction}---Many fundamental problems in quantum information theory, such as state discrimination~\cite{Helstrom1969,Holevo1972}, channel capacities~\cite{HolevoWerner2001,Wang2018semidefinite}, entanglement quantification~\cite{Rains2001SDP}, and quantum hypothesis testing, can be cast as semidefinite programs (SDPs)~\cite{Mironowicz2024SDP,Watrous2018}. Efficient SDP solvers are therefore a basic tool for quantum information research~\cite{vandenberghe1996semidefinite,NesterovNemirovski1994,Arora2012MMW}, with quantum-native solvers~\cite{BrandaoSvore2017,Brandao2019QuantumSDP,vanApeldoorn2019SDP,vanApeldoornGilyen2019,KerenidisPrakash2020QIPM} especially attractive when the SDP dimension grows exponentially with the underlying system size. In each of these contexts, the SDP variable belongs to a different set, dictated by which constituent of quantum mechanics is being optimized: density operators $\rho\in\DD$, where $\DD$ is the set of $d \times d$ density operators, for states; measurement operators $0\preceq M\preceq I$, where $\MM$ is the set of $d \times d$ measurement operators, for POVM elements; and the unbounded positive semidefinite cone $\PP \coloneqq \{X \succeq 0\}$ for unnormalized observables. The eigenvalues of operators in these three sets obey precisely the three rules of the corresponding quantum statistics~\cite{Pathria2011Statistical}: trace-one normalization for Boltzmann, the Pauli exclusion principle $0\le m_{j}\le 1$ for Fermi--Dirac, and unrestricted occupation numbers $x_{j}\ge 0$ for Bose--Einstein.

Drawing on this observation, our line of work develops thermodynamic frameworks for SDPs across these three distributions~\cite{liu2025sdp,liu2026sdp_fermi,minervini2026sdp_bose}. Adopting Jaynes' statistical-mechanics mindset~\cite{Jaynes1957}, in each framework the linear SDP objective is interpreted as a system energy, and the optimization is recast as a free-energy minimization at a strictly positive temperature $T>0$, by regularizing the energy with a statistics-matched quantum entropy scaled by $T$. In Refs.~\cite{liu2026sdp_fermi,minervini2026sdp_bose}, however, the fermions and bosons entered only as an effective, mode-by-mode description of the spectrum of the SDP variable, and the three frameworks were developed as three parallel constructions.

In this Letter, we show that the correspondence is not an analogy but an identity, and that the three constructions can be unified. Our main result (Theorem~\ref{thm:fock}) states that the Fermi--Dirac and Bose--Einstein regularized SDPs are equal to the constrained free-energy minimization of a genuine many-body system, an ideal Fermi or Bose gas on the Fock space over the $d$-dimensional Hilbert space of the SDP, whose Hamiltonian and charges are the second quantizations $\dG(H)$ and $\dG(Q_{i})$ of the SDP data and whose one-body density matrix is the SDP variable; the Boltzmann framework corresponds to the case of a single particle. The Fermi--Dirac and Bose--Einstein frameworks therefore correspond to the Boltzmann framework of Ref.~\cite{liu2025sdp} applied to two many-body systems. Consequently, the features that the three frameworks were observed to share, namely an unconstrained concave dual in a chemical-potential vector, a thermal operator as primal optimum, and dual derivatives that are thermal expectation values, all follow from Ref.~\cite{liu2025sdp}; in particular, the three dual objectives are the same function, $\mu\cdot q-T\ln Z(\mu)$, of the grand partition function $Z$ of the corresponding system. The identification extends to the divergences: the Fermi--Dirac and Bose--Einstein relative entropies of Refs.~\cite{liu2026sdp_fermi,minervini2026sdp_bose} are the Umegaki relative entropies between quasi-free (Gaussian) states~\cite{Dierckx2008,LewinSabin2014}, so that their monotonicity under gauge-covariant fermionic and bosonic Gaussian channels follows from the data-processing inequality; for bosons, this recovers the monotonicity under attenuator, amplifier, and additive-noise channels proved in Ref.~\cite{minervini2026sdp_bose} and extends it to general multimode channels. We first recall the common formulation of the three frameworks, then state the many-body equivalence and its consequences (with proofs in the Supplemental Material, Appendices~\hyperref[app:A]{A}--\hyperref[app:C]{C}), and finally review the results of Refs.~\cite{liu2025sdp,liu2026sdp_fermi,minervini2026sdp_bose} in this light.

\textit{Unified thermodynamic framework}---A general SDP is specified by a tuple of Hermitian operators, $\mathcal{Q}\coloneqq (H,Q_{1},\dots,Q_{c})$, each acting on a $d$-dimensional Hilbert space $\Hil$ and a tuple of constraint values, $q\coloneqq (q_{1},\dots,q_{c})\in\mathbb{R}^{c}$. The optimization variable $X$ is constrained to a feasible set $\mathcal{C}_{\stat}\in\{\DD,\MM,\PP\}$ matching one of the three quantum sets, where the label $\stat\in\{\mathrm{B},\mathrm{FD},\mathrm{BE}\}$ runs over the three kinds of statistics, namely Boltzmann (B), Fermi--Dirac (FD), and Bose--Einstein (BE). Also, $\DD$ denotes the set of $d \times d$ density operators ($\rho\succeq 0$, $\Tr[\rho]=1$), $\MM$ denotes the set of $d \times d$ measurement operators ($0\preceq M\preceq I$), and $\PP \coloneqq \{X\succeq 0\}$ denotes the unbounded positive semidefinite cone. The SDP optimizes the linear objective $\Tr[HX]$ over $X\in\mathcal{C}_{\stat}$, subject to the affine constraints $\Tr[Q_{i}X]=q_{i}$ for all $i\in[c]\coloneqq\{1,\dots,c\}$:
\begin{equation}\label{eq:sdp}
E^{\stat}(\mathcal{Q},q)\coloneqq \min_{\substack{X\in\mathcal{C}_{\stat},\\ \Tr[Q_{i}X]=q_{i}\ \forall i\in[c]}}\Tr[HX].
\end{equation}
Physically, $H$ is a Hamiltonian, each $Q_{i}$ is a non-commuting charge, and the eigenvalues of $X$ play the role of expected occupation numbers under the matched statistics; Theorem~\ref{thm:fock} below makes this identification literal.

Inspired by the fact that physical systems operate at a strictly positive temperature, we minimize the free energy rather than the energy, obtaining a smooth approximation to the original SDP:
\begin{equation}\label{eq:freeen}
F_{T}^{\stat}(\mathcal{Q},q)\coloneqq \min_{\substack{X\in\mathcal{C}_{\stat}\\ \Tr[Q_{i}X]=q_{i}\ \forall i\in[c]}}\left\{\Tr[HX]-T\,S_{\stat}(X)\right\}
\end{equation}
at temperature $T>0$. Here $S_{\stat}$ is the entropy matched to the statistics $\stat$: the von Neumann entropy for the trace-normalized set $\DD$, the Fermi--Dirac entropy for the operator-bounded set $\MM$, and the Bose--Einstein entropy for the unbounded cone $\PP$ (see Table~\ref{tab:three-frameworks} for their explicit expressions).

\begin{table*}[t]
\centering
\renewcommand{\arraystretch}{1.35}
\setlength{\tabcolsep}{3pt}
\footnotesize
\begin{tabular}{p{0.14\textwidth} p{0.18\textwidth} p{0.325\textwidth} p{0.305\textwidth}}
\hline\hline
 & \textbf{Boltzmann}~\cite{liu2025sdp} & \textbf{Fermi--Dirac}~\cite{liu2026sdp_fermi} & \textbf{Bose--Einstein}~\cite{minervini2026sdp_bose} \\
\hline
Set $\mathcal{C}_{\stat}$ & $\rho\succeq 0$,\ $\Tr[\rho]=1$ & $0\preceq M\preceq I$ & $X\succeq 0$ \\
Eigenvalue rule & trace-one probabilities & Pauli exclusion: $0\le m_{j}\le 1$ & unrestricted occupations: $x_{j}\ge 0$ \\
Many-body system & single particle on $\Hil$ & ideal Fermi gas on $\Fock_{-}(\Hil)$ & ideal Bose gas on $\Fock_{+}(\Hil)$ \\
Entropy & $S(\rho)\coloneqq -\Tr[\rho\ln\rho]$ & $S_{\mathrm{FD}}(M)\coloneqq-\Tr[M\ln M+(I-M)\ln(I-M)]$ & $S_{\mathrm{BE}}(X)\coloneqq-\Tr[X\ln X - (X+I)\ln(X+I)]$ \\
Relative entropy & $D(\rho\|\sigma)\coloneqq\Tr[\rho(\ln\rho-\ln\sigma)]$ & $D_{\mathrm{FD}}(M\|N)\coloneqq D(M\|N)+D(I-M\|I-N)$ & $D_{\mathrm{BE}}(X\|Y)\coloneqq D(X\|Y)-D(X+I\|Y+I)$ \\
Partition function & $Z_{\mathrm{B}}(\mu)\coloneqq\Tr[e^{-\Km/T}]$ & $Z_{\mathrm{FD}}(\mu)\coloneqq\det(I+e^{-\Km/T})$ & $Z_{\mathrm{BE}}(\mu)\coloneqq\det(I-e^{-\Km/T})^{-1}$ \\
Primal optimum & $\rho_{T}(\mu)\coloneqq \dfrac{e^{-\Km/T}}{Z_{\mathrm{B}}(\mu)}$ & $M_{T}(\mu)\coloneqq(e^{\Km/T}+I)^{-1}$ & $X_{T}(\mu)\coloneqq(e^{\Km/T}-I)^{-1}$ \\
Optimum's name & Boltzmann (Gibbs) state & Fermi--Dirac thermal measurement & Bose--Einstein thermal operator \\
Target SDP & trace-normalized & operator-bounded & standard form; unbounded PSD cone \\
\hline\hline
\end{tabular}
\caption{The three thermodynamic frameworks are three instances of one construction (Theorem~\ref{thm:fock}): the constrained free-energy minimization of a quantum system with Hamiltonian $\dG(H)$ and charges $\dG(Q_{i})$ on the many-body space of the third row, whose one-body density matrix is the SDP variable and whose grand canonical Gibbs state has the primal optimum as one-body density matrix. In all three cases the dual objective is $f_{T}^{\stat}(\mu)=\mu\cdot q-T\ln Z_{\stat}(\mu)$, with $\Km = H - \mu\cdot Q$ the grand canonical Hamiltonian, and the relative entropy is the Bregman divergence generated by $-S_{\stat}$, equal to the Umegaki relative entropy between quasi-free states (Proposition~\ref{prop:relent}). The frameworks differ in the feasible set, the entropy, the thermal-operator form, and the quantum primitives discussed in the main text.}
\label{tab:three-frameworks}
\end{table*}

Lagrangian duality~\cite{Boyd2004convex}, combined with the quantum relative entropy adapted to each set (Umegaki for the trace-normalized set, Fermi--Dirac for the operator-bounded case, Bose--Einstein for the unbounded one; see Table~\ref{tab:three-frameworks} for the expressions), yields an unconstrained concave dual
\begin{equation}\label{eq:dual}
F_{T}^{\stat}(\mathcal{Q},q)=\sup_{\mu\in\mathbb{R}^{c}}f_{T}^{\stat}(\mu),
\end{equation}
whose Lagrange multipliers $\mu$ play the role of chemical potentials conjugate to the conserved charges $Q_{i}$. The unique primal optimum is a thermal operator $X_{T}^{\stat}(\mu^{\star})$ of the matched statistics, parametrized by the chemical-potential vector $\mu$ through the grand canonical Hamiltonian $\Km\coloneqq H-\mu\cdot Q$; see the ``Primal optimum'' row of Table~\ref{tab:three-frameworks} for the explicit form in each framework. In all three frameworks, the dual derivatives admit closed forms as thermal expectation values:
\begin{equation}\label{eq:grad}
\frac{\partial f_{T}^{\stat}}{\partial\mu_{i}}=q_{i}-\Tr\!\left[X_{T}^{\stat}(\mu)\,Q_{i}\right],
\end{equation}
and the Hessian is a thermal covariance of pairs of charges (see Eq.~\eqref{eq:hessian} below). Together with the concavity of $f_{T}^{\stat}$, these closed-form derivatives guarantee that gradient ascent and Newton's method converge globally to the dual optimum~\cite{bubeck2015convex}. Each step of the optimization is thus, physically, the measurement of the charges $\mathcal{Q}$ on the thermal operator formed from $\Km/T$. As $T\to 0$, the limit $F_{T}^{\stat}\to E^{\stat}$ holds, and the temperature $T$ is our control knob, smoothly trading the precision of the approximation to the original SDP against the conditioning of the dual problem.

\textit{Second quantization: one framework, three systems}---Let $\Fock_{-}(\Hil)$ and $\Fock_{+}(\Hil)$ denote the fermionic and bosonic Fock spaces over $\Hil\simeq\mathbb{C}^{d}$, respectively, with creation and annihilation operators $a_{j}^{\dagger}$ and $a_{j}$ attached to an orthonormal basis $\{|j\rangle\}_{j\in[d]}$ of $\Hil$ and obeying the canonical anticommutation or commutation relations~\cite{Bratteli1997}. The second quantization of a Hermitian operator $A$ on $\Hil$ is the one-body operator
\begin{equation}\label{eq:dGamma}
\dG(A)\coloneqq\sum_{j,k\in[d]}\langle j|A|k\rangle\,a_{j}^{\dagger}a_{k},
\end{equation}
which is linear in $A$ and independent of the basis; e.g., $\dG(I)=\Nhat$ is the particle-number operator. For a state~$\omega$, i.e., a density operator on the Fock space, and an operator $X$ on the Fock space, we write $\omega(X)\coloneqq\Tr[ X \omega]$ for the expectation value of $X$ with respect to $\omega$. Every state $\omega$ on $\Fock_{\pm}(\Hil)$ (with finite mean particle number, for bosons) has a one-body density matrix $\gamma_{\omega}$ on $\Hil$, the $d\times d$ matrix of expectation values $\langle j|\gamma_{\omega}|k\rangle\coloneqq\omega(a_{k}^{\dagger}a_{j})$: its diagonal entries correspond to the mean occupation numbers of the modes $|j\rangle$, and its off-diagonal entries correspond to the coherences between modes. It satisfies
\begin{equation}\label{eq:1rdm}
\Tr[A\gamma_{\omega}]=\omega\big(\dG(A)\big)\quad\text{for every Hermitian }A,
\end{equation}
so that $\Tr[\gamma_{\omega}]=\omega(\Nhat)$ is equal to the mean particle number, and $\gamma_{\omega}$ is a density operator only for a single particle. More generally, for a unit vector $\varphi\in\Hil$, $\langle\varphi|\gamma_{\omega}|\varphi\rangle=\omega(a(\varphi)^{\dagger}a(\varphi))$ is the mean number of particles in the single-particle state $\varphi$, where $a(\varphi)\coloneqq\sum_{j}\langle\varphi|j\rangle a_{j}$ annihilates a particle in $\varphi$. This number is nonnegative, because $a(\varphi)^{\dagger}a(\varphi)\succeq0$, so that $\gamma_{\omega}\succeq0$; for fermions, the canonical anticommutation relations (i.e., $\{a_j, a_k^\dag\}=\delta_{jk}$) make $a(\varphi)^{\dagger}a(\varphi)$ a projector, i.e., the mode $\varphi$ holds at most one particle (Pauli exclusion), so that also $\gamma_{\omega}\preceq I$. These are precisely the constraints of Table~\ref{tab:three-frameworks}: $0\preceq\gamma_{\omega}\preceq I$ for fermions and $\gamma_{\omega}\succeq0$ for bosons (Lemma~\ref{lem:constraints} in \appref{A})~\cite{Coleman1963}. For a single particle there is no exchange statistics: its state space is $\Hil$ itself, $\dG(A)$ acts as $A$, and $\gamma_{\omega}=\omega$ is a density operator. Accordingly, we set $\Fock_{\mathrm{B}}\coloneqq\Hil$ with $\dG_{\mathrm{B}}(A)\coloneqq A$, $\Fock_{\mathrm{FD}}\coloneqq\Fock_{-}(\Hil)$, and $\Fock_{\mathrm{BE}}\coloneqq\Fock_{+}(\Hil)$, with $\dG_{\mathrm{FD}}=\dG_{\mathrm{BE}}\coloneqq\dG$.

\begin{theorem}\label{thm:fock}
For every $\stat\in\{\mathrm{B},\mathrm{FD},\mathrm{BE}\}$ and $T>0$,
\begin{multline}
F_{T}^{\stat}(\mathcal{Q},q)=\min_{\omega}\Big\{\,\omega\big(\dG_{\stat}(H)\big)-T\,S(\omega)\ :\ \\
\omega\big(\dG_{\stat}(Q_{i})\big)=q_{i}\ \ \forall i\in[c]\Big\},\label{eq:fock-free-energy}
\end{multline}
where the minimum is over every state $\omega$ on $\Fock_{\stat}$ and $S$ is the von Neumann entropy. The minimizer is unique and equal to the grand canonical Gibbs state $\omega_{T}^{\stat}(\mu^{\star})$, where
\begin{equation}\label{eq:gibbs-fock}
\omega_{T}^{\stat}(\mu)\coloneqq\frac{e^{-\dG_{\stat}(\Km)/T}}{Z_{\stat}(\mu)},\quad Z_{\stat}(\mu)\coloneqq\Tr_{\Fock_{\stat}}\!\big[e^{-\dG_{\stat}(\Km)/T}\big],
\end{equation}
the one-body density matrix of $\omega_{T}^{\stat}(\mu^{\star})$ is the primal optimum $X_{T}^{\stat}(\mu^{\star})$ of Table~\ref{tab:three-frameworks}, and the dual objective in~\eqref{eq:dual} is
\begin{equation}\label{eq:dual-Z}
f_{T}^{\stat}(\mu)=\mu\cdot q-T\ln Z_{\stat}(\mu),
\end{equation}
for all $\mu\in\mathbb{R}^{c}$ if $\stat\in\{\mathrm{B},\mathrm{FD}\}$ and for all $\mu$ with $\Km\succ0$ if $\stat=\mathrm{BE}$.
\end{theorem}

Theorem~\ref{thm:fock} says that the Fermi--Dirac and Bose--Einstein frameworks are the Boltzmann framework of Ref.~\cite{liu2025sdp} applied to an ideal Fermi or Bose gas whose single-particle space carries the SDP: the Hamiltonian is $\dG(H)$, the charges are $\dG(Q_{i})$, the SDP variable is the one-body density matrix of the gas, the thermal operators of Table~\ref{tab:three-frameworks} are those of its Gibbs states, and $Z_{\stat}$ is its grand partition function, $Z_{\mathrm{FD}}\coloneqq \det(I+e^{-\Km/T})$ and $Z_{\mathrm{BE}}\coloneqq \det(I-e^{-\Km/T})^{-1}$; the latter requires $\Km\succ0$, the dual-feasible set of Ref.~\cite{minervini2026sdp_bose}, at whose boundary the Bose gas condenses and $Z_{\mathrm{BE}}$ diverges. Inserting these expressions into~\eqref{eq:dual-Z} recovers the dual objectives derived separately in Refs.~\cite{liu2025sdp,liu2026sdp_fermi,minervini2026sdp_bose}. In thermodynamic terms, $-T\ln Z_{\stat}(\mu)$ is the grand potential of the system~\cite{Pathria2011Statistical}, so that the dual objective~\eqref{eq:dual-Z} is this grand potential plus the term $\mu\cdot q$, which is linear in the chemical potentials.

The proof of Theorem~\ref{thm:fock}, given in \appref{B}, rests on two facts. First, $\gamma_{\omega}\in\mathcal{C}_{\stat}$ for every state $\omega$, so that, by~\eqref{eq:1rdm}, a feasible many-body state yields a feasible SDP variable with the same energy and charges. Second, among all states with a given one-body density matrix $\gamma$ in the interior of $\mathcal{C}_{\stat}$, the gauge-invariant quasi-free (Gaussian) state
\begin{equation}\label{eq:quasifree}
\omega_{\gamma}\propto e^{-\dG(\Lambda_{\gamma})},\qquad \Lambda_{\gamma}\coloneqq\begin{cases}\ln(\gamma^{-1}-I), & \stat=\mathrm{FD},\\ \ln(\gamma^{-1}+I), & \stat=\mathrm{BE},\end{cases}
\end{equation}
has the largest von Neumann entropy, and this entropy equals $S_{\stat}(\gamma)$, the Fermi--Dirac or Bose--Einstein entropy of Eq.~\eqref{eq:freeen} and Table~\ref{tab:three-frameworks}, evaluated at $\gamma$~\cite{BachLiebSolovej1994,WolfGiedkeCirac2006}. Here, $\Lambda_{\gamma}$ is the one-body Hamiltonian, in units of the temperature, whose Gibbs state $\omega_{\gamma}$ has one-body density matrix $\gamma$; it is obtained by inverting the Fermi--Dirac and Bose--Einstein distributions, $\gamma=(e^{\Lambda_{\gamma}}+I)^{-1}$ for fermions and $\gamma=(e^{\Lambda_{\gamma}}-I)^{-1}$ for bosons, so that, e.g., $\Lambda_{\gamma}=\Km/T$ for $\gamma=X_{T}^{\stat}(\mu)$. More precisely, $S(\omega)=S_{\stat}(\gamma_{\omega})-D(\omega\|\omega_{\gamma_{\omega}})$ for every state $\omega$ with $\gamma_{\omega}$ in the interior of $\mathcal{C}_{\stat}$, where $D(\omega\|\omega')\coloneqq\Tr[\omega(\ln\omega-\ln\omega')]\ge0$ is the Umegaki relative entropy of Table~\ref{tab:three-frameworks}, here evaluated on states of the Fock space. This holds because $\ln\omega_{\gamma}$ is affine in the one-body operators and the entropy of $\omega_{\gamma}$ is a sum of single-mode thermal entropies. Minimizing in~\eqref{eq:fock-free-energy} first at fixed one-body density matrix and then over the latter yields~\eqref{eq:freeen}; the Gibbs state~\eqref{eq:gibbs-fock} is the quasi-free state whose one-body density matrix is the SDP optimum; and~\eqref{eq:dual-Z} is the dual of Ref.~\cite{liu2025sdp} for the many-body system, because $\dG(H)-\sum_{i}\mu_{i}\dG(Q_{i})=\dG(\Km)$.

For the Hessian, we use a formula due to Jaynes~\cite{Jaynes1962} and recalled in Ref.~\cite{liu2025sdp}: the Hessian of the logarithm $\ln Z$ of a grand partition function with respect to the chemical potentials equals $1/T^{2}$ times the (Kubo--Mori) covariance matrix of the conserved charges in the corresponding Gibbs state. Applied to the many-body system of Theorem~\ref{thm:fock}, whose charges are $\dG_{\stat}(Q_{i})$, it gives, for all three statistics (see \appref{B} for a derivation),
\begin{equation}\label{eq:hessian}
\frac{\partial^{2}f_{T}^{\stat}}{\partial\mu_{i}\partial\mu_{j}}=-\frac{1}{T}\Cov^{\mathrm{KM}}_{\omega_{T}^{\stat}(\mu)}\!\big(\dG_{\stat}(Q_{i}),\dG_{\stat}(Q_{j})\big),
\end{equation}
with $\Cov^{\mathrm{KM}}_{\omega}(A,B)\coloneqq\int_{0}^{1}ds\,\Tr[\omega^{s}A\,\omega^{1-s}B]-\omega(A)\omega(B)$ the Kubo--Mori covariance. Because $f_{T}^{\stat}$ is the same function whether it is written on $\Hil$ or on the Fock space, rewriting~\eqref{eq:hessian} in terms of one-body operators on $\Hil$ reproduces the Hessian formulas of Refs.~\cite{liu2026sdp_fermi,minervini2026sdp_bose}, and Newton's method is, in all three frameworks, natural-gradient ascent in the Kubo--Mori geometry of the many-body Gibbs family~\cite{liu2025sdp,Patel2025a}.

The identification extends to the divergences of Table~\ref{tab:three-frameworks}.

\begin{proposition}\label{prop:relent}
Let $\gamma,\sigma$ lie in the interior of $\mathcal{C}_{\stat}$, and let $\omega$ be a state on $\Fock_{\stat}$ with $\gamma_{\omega}=\gamma$. Then
\begin{equation}\label{eq:pythagoras}
D(\omega\|\omega_{\sigma})=D(\omega\|\omega_{\gamma})+D_{\stat}(\gamma\|\sigma).
\end{equation}
In particular, $D(\omega_{\gamma}\|\omega_{\sigma})=D_{\stat}(\gamma\|\sigma)$ and $D_{\stat}(\gamma\|\sigma)=\min\{D(\omega\|\omega_{\sigma}):\gamma_{\omega}=\gamma\}$. 
\end{proposition}

For quasi-free states, the identity $D(\omega_{\gamma}\|\omega_{\sigma})=D_{\stat}(\gamma\|\sigma)$ is known in mathematical physics~\cite{Dierckx2008,LewinSabin2014}; Proposition~\ref{prop:relent} connects it with the divergences of Refs.~\cite{liu2026sdp_fermi,minervini2026sdp_bose}. Equation~\eqref{eq:pythagoras}, which extends it to every state $\omega$, is a Pythagorean theorem for the relative entropy, in the sense of information geometry~\cite{Csiszar1975,AmariNagaoka2000}. The states $\omega$ with $\gamma_{\omega}=\gamma$ form a linear family, defined by the linear constraints $\omega(a_{k}^{\dagger}a_{j})=\langle j|\gamma|k\rangle$, whereas the quasi-free states form an exponential family; the two families meet at $\omega_{\gamma}$, which, by Proposition~\ref{prop:relent}, is the relative-entropy projection of $\omega_{\sigma}$ onto the linear family. Just as the squared hypotenuse of a right triangle is the sum of the squared legs, for the triangle with vertices $\omega$, $\omega_{\gamma}$, and $\omega_{\sigma}$ and right angle at $\omega_{\gamma}$, the relative entropy $D(\omega\|\omega_{\sigma})$ splits into the relative entropy $D(\omega\|\omega_{\gamma})$ within the linear family and the relative entropy $D(\omega_{\gamma}\|\omega_{\sigma})=D_{\stat}(\gamma\|\sigma)$ within the exponential family.

By Proposition~\ref{prop:relent}, $D_{\mathrm{FD}}$ and $D_{\mathrm{BE}}$ are Umegaki relative entropies between quasi-free states, so that their natural channels are channels on Fock space that preserve quasi-free states. A channel on $\Hil$, by contrast, is in general not a physical operation on the gas, which explains why $D_{\mathrm{BE}}$ violates the data-processing inequality for channels on $\Hil$~\cite{minervini2026sdp_bose}, even for mixed-unitary ones (see \appref{C}). Together with the data-processing inequality for the Umegaki relative entropy~\cite{Lindblad1975}, this identification yields the following.

\begin{proposition}\label{prop:dpi}
Let $\mathcal{N}$ be a channel on $\Fock_{\stat}$ mapping every quasi-free state $\omega_{\gamma}$ to the quasi-free state $\omega_{A\gamma A^{\dagger}+B}$, for fixed operators $A,B$ on $\Hil$; such gauge-covariant quasi-free channels~\cite{Caruso2008,GiovannettiHolevoGarciaPatron2015,Dierckx2008} arise, e.g., from a gauge-covariant linear coupling of the system modes to environment modes prepared in a quasi-free state, such as passive mixing or, for bosons, phase-insensitive amplification, followed by tracing out the environment. Then, for all $\gamma,\sigma$ in the interior of $\mathcal{C}_{\stat}$,
\begin{equation}\label{eq:dpi}
D_{\stat}(\gamma\|\sigma) \geq D_{\stat}\big(A\gamma A^{\dagger}+B\,\big\|\,A\sigma A^{\dagger}+B\big) .
\end{equation}
\end{proposition}

For bosons, $A=\sqrt{\kappa}\,I$ and $B=\nu I$, with $(\kappa,\nu)$ the parameters denoted $(a,b)$ in Ref.~\cite{minervini2026sdp_bose}, give the attenuator, amplifier, and additive-noise channels acting identically on all modes, so that Proposition~\ref{prop:dpi} recovers the monotonicity of $D_{\mathrm{BE}}$ under these channels, proved in Ref.~\cite{minervini2026sdp_bose} through a scalar integral representation. Proposition~\ref{prop:dpi} also extends it to general multimode gauge-covariant bosonic Gaussian channels, for which $B\succeq0$ and $B\succeq AA^{\dagger}-I$~\cite{Caruso2008,GiovannettiHolevoGarciaPatron2015}, and yields the analogous monotonicity of $D_{\mathrm{FD}}$ under fermionic quasi-free channels~\cite{Dierckx2008}, e.g., those obtained by passive linear mixing with environment modes, for which $0\preceq B\preceq I-AA^{\dagger}$.

The Fermi--Dirac divergence has two further properties that its bosonic counterpart lacks (see \appref{C} for proofs). Because $D_{\mathrm{FD}}(\gamma\|\sigma)=D(\gamma\|\sigma)+D(I-\gamma\|I-\sigma)$ is a sum of Umegaki relative entropies whose arguments are affine in $(\gamma,\sigma)$, it is jointly convex and, since $\Phi(I-\gamma)=I-\Phi(\gamma)$ for every unital channel $\Phi$ on $\Hil$, it satisfies $D_{\mathrm{FD}}(\Phi(\gamma)\|\Phi(\sigma))\le D_{\mathrm{FD}}(\gamma\|\sigma)$; non-unital channels need not preserve the set $0\preceq\gamma\preceq I$, so that no such statement is expected for them. By contrast, $D_{\mathrm{BE}}(\gamma\|\sigma)=D(\gamma\|\sigma)-D(\gamma+I\|\sigma+I)$ is a difference of such terms and is not jointly convex~\cite{minervini2026sdp_bose}, so that, for $D_{\mathrm{BE}}$, Proposition~\ref{prop:dpi} is the relevant form of data processing. In the Boltzmann case, $\Fock_{\mathrm{B}}=\Hil$, every state is its own quasi-free state, and $D$ satisfies the data-processing inequality for all channels on $\Hil$~\cite{Lindblad1975}.

\textit{The three systems, side by side}---Theorem~\ref{thm:fock} assigns a physical system to each framework. We now revisit the three frameworks in this light, summarizing the results of Refs.~\cite{liu2025sdp,liu2026sdp_fermi,minervini2026sdp_bose} that Table~\ref{tab:three-frameworks} condenses.

\textit{Boltzmann framework}~\cite{liu2025sdp}: a single particle. The SDP variable is the state $\rho$ of one particle (trace-one normalization), and the primal optimum $\rho_{T}(\mu)$ is a non-Abelian grand canonical thermal state~\cite{YungerHalpern2016,YungerHalpern2020}, that is, a quantum Boltzmann machine~\cite{Amin2018,Kieferova2017}. The framework natively handles SDPs whose variable is a quantum state; its constrained free-energy minimization is precisely the training step of a quantum Boltzmann machine on a target observable, with the chemical potentials $\mu$ as parameters~\cite{Amin2018,Kieferova2017,Patel2025a,patel2025quantumboltzmannmachinelearning}, and it has been validated numerically~\cite{minervini2025constrained}. A simple reduction extends it to general SDPs whenever an a priori upper bound on the optimal trace of the primal variable can be guessed.

\textit{Fermi--Dirac framework}~\cite{liu2026sdp_fermi}: an ideal Fermi gas. The eigenvalues $0\le m_{j}\le 1$ of $M$ are the mean occupation numbers of its modes (Pauli exclusion), and the primal optimum $M_{T}(\mu)$, the one-body density matrix of the gas at temperature $T$, is a Fermi--Dirac thermal measurement. The framework handles quantum hypothesis testing problems (symmetric, asymmetric, and composite binary hypothesis testing, and binary classification) as special cases: as $T\to0$ the Fermi--Dirac thermal measurement approaches the Helstrom--Holevo optimal measurement~\cite{Helstrom1969,Holevo1972}, which, for symmetric hypothesis testing between $\rho$ and $\sigma$, is in this language the Fermi sea of the effective single-particle Hamiltonian $\sigma-\rho$, with every mode of negative energy occupied. The framework also introduces \emph{Fermi--Dirac machines}, a quantum machine learning paradigm complementary to quantum Boltzmann machines, in which the trained object is a parametrized Fermi--Dirac thermal measurement rather than a parametrized state.

\textit{Bose--Einstein framework}~\cite{minervini2026sdp_bose}: an ideal Bose gas. The eigenvalues $x_{j}\ge 0$ of $X$ are mean occupation numbers with no exclusion principle, which is the natural setting of SDPs in standard form, whose variable is an unbounded positive semidefinite operator with no a priori bound on its trace; the primal optimum $X_{T}(\mu)$ is the one-body density matrix of the gas, a Bose--Einstein thermal operator. Dual feasibility, $\Km\succ0$, is precisely the condition for the gas to admit a Gibbs state, and primal-feasible operators of arbitrarily large trace correspond to the onset of Bose--Einstein condensation. The approximation error of the regularized SDP can be made independent of the Hilbert-space dimension, depending instead on the ground-space degeneracy and the spectral gap of $K_{\mu^{\star}}$. The framework also introduces the Bose--Einstein relative entropy $D_{\mathrm{BE}}$, a Bregman divergence~\cite{Tsuda2005matrix,Dhillon2007Matrix} on the unbounded positive semidefinite cone, which Proposition~\ref{prop:relent} identifies with the relative entropy between bosonic quasi-free states.

\textit{Hybrid quantum--classical algorithms}---In the language of Theorem~\ref{thm:fock}, each gradient component~\eqref{eq:grad} is the expectation of the one-body observable $\dG(Q_{i})$ in the quasi-free Gibbs state of the gas. Because quasi-free states are determined by their one-body density matrices, this expectation reduces to $\Tr[Q_{i}X_{T}^{\stat}(\mu)]$ on the $d$-dimensional space $\Hil$: the many-body Gibbs state never has to be prepared, and the optimization reduces, in all three frameworks, to estimating thermal expectation values of the input observables on the thermal operator of $\Km/T$. In all three settings the hybrid quantum--classical estimator combines the same three ingredients: \emph{(i)} \emph{Hamiltonian simulation}~\cite{Low2019hamiltonian} of $e^{-i\Km t}$ for an evolution time $t$; \emph{(ii)} the \emph{Hadamard test}, which extracts the relevant trace from the simulated dynamics; and \emph{(iii)} \emph{classical random sampling} of certain parameters of the estimator. The framework-specific elements complete the recipe: the Boltzmann framework additionally calls a Gibbs-state preparation subroutine to prepare $\rho_{T}(\mu)$, on which Hamiltonian simulation, Hadamard tests, and a high-peak-tent random sampling of evolution times then realize the gradient and Hessian estimators~\cite{liu2025sdp}; the Fermi--Dirac framework realizes the Fermi--Dirac thermal measurement directly on a hybrid qubit--qumode register via Schr\"odingerization \cite{Jin2023,Jin2024} and the power of one qumode \cite{Liu2016}, then combines this primitive with Hamiltonian simulation and Hadamard tests~\cite{liu2026sdp_fermi}; and the Bose--Einstein framework expands the thermal operator as a geometric series and converts each term by sampling the evolution time $t$ from a Cauchy distribution, with Hadamard tests extracting the relevant traces~\cite{minervini2026sdp_bose}.

Concavity of the dual objective and a bounded smoothness parameter $L_{T}$ guarantee, in all three frameworks, that stochastic gradient ascent~\cite{Robbins1951stochastic,bubeck2015convex} reaches an $\eps$-suboptimal dual point in $\mathcal{O}(L_{T}\|\mu_{T}^{\star}\|^{2}/\eps)$ iterations, with second-order Newton-type variants available throughout. Thus the end-to-end runtimes are polynomial in the inverse target precision $1/\eps$ and in the number $c$ of constraints, with the remaining polynomial dependencies being framework-specific. Together, these primitives realize hybrid quantum--classical SDP solvers in each of the three frameworks.

\textit{Highlights and applications}---The three frameworks share their thermodynamic structure but offer complementary strengths, dictated by the SDP at hand.

\textit{(B) Boltzmann.}
The Boltzmann framework is the natural choice when the SDP variable is a quantum state. It provides the language for quantum Boltzmann machine learning~\cite{Patel2025a,patel2025quantumboltzmannmachinelearning}, where the model state to be trained is itself a parametrized thermal state and the SDP solved here is the training step, and it yields a hybrid quantum--classical SDP solver that, in a regime where the dual variable has large norm, achieves a polynomial improvement over existing quantum SDP solvers~\cite{vanApeldoorn2019SDP}.

\textit{(FD) Fermi--Dirac.}
The Fermi--Dirac framework is the natural choice for measurement-bounded SDPs. Quantum hypothesis testing maps onto it directly, and Fermi--Dirac thermal measurements provide a smooth, gradient-trainable substitute for the sharp Helstrom--Holevo measurement, with provable closeness at low temperature. The same primitive underlies a recently proposed soft principal-component analysis with a one-shot quantum calibration~\cite{yuan2026qPCA}, a thermal counterpart to quantum principal-component analysis~\cite{Lloyd2014qPCA}, and Fermi--Dirac machines have recently been reinterpreted as quantizations of classical neurons~\cite{he2026fermidirac,he2026canonical}, with complexity-theoretic evidence that they can learn functions beyond classical reach.

\textit{(BE) Bose--Einstein.}
The Bose--Einstein framework targets SDPs in standard form. Because the cone $X\succeq 0$ is unbounded, no a priori upper bound on the trace of the primal variable is required, in contrast with other quantum SDP solvers~\cite{vanApeldoorn2019SDP} and with the Boltzmann reduction. When the grand canonical Hamiltonian at the optimum is gapped, the approximation error of the regularized SDP and the resulting runtime scale with the ground-space degeneracy and the spectral gap rather than with the Hilbert-space dimension, improving over classical interior-point methods, whose duality gap on the central path scales linearly with $d$~\cite{NesterovNemirovski1994}. Finally, just as the von Neumann and Fermi--Dirac entropies give rise to quantum Boltzmann machines and Fermi--Dirac machines, the Bose--Einstein regularization suggests \emph{Bose--Einstein machines}, in which the trained object is a parametrized Bose--Einstein thermal operator.

\textit{Conclusion}---Theorem~\ref{thm:fock} replaces three parallel thermodynamic analogies by one statement: an SDP regularized by the entropy matched to its constraint set is the constrained free-energy minimization of a quantum system with Hamiltonian $\dG(H)$ and charges $\dG(Q_{i})$, namely a single particle, an ideal Fermi gas, or an ideal Bose gas, and the SDP variable is the one-body density matrix of that system. The concave dual, its gradient and Hessian, and the resulting hybrid quantum--classical solvers are those of the Boltzmann framework applied to this system, whereas the one-body reductions, error bounds, and quantum primitives of Refs.~\cite{liu2026sdp_fermi,minervini2026sdp_bose} are what make them efficient. The Fermi--Dirac and Bose--Einstein relative entropies are Umegaki relative entropies between quasi-free states and inherit data processing under gauge-covariant Gaussian channels. In this language, quantum Boltzmann machines train the state of a single particle, while Fermi--Dirac and Bose--Einstein machines train the one-body density matrix of an ideal gas; whether interacting gases, whose energy involves two-body reduced density matrices, lead to a useful hierarchy of optimization problems beyond SDPs is an interesting open question.
\medskip

\begin{acknowledgments}
NL acknowledges funding from the Science and Technology Commission of Shanghai Municipality (STCSM) grant no.~24LZ1401200 (21JC1402900), NSFC grants No.~12471411 and No.~12341104, the Shanghai Jiao Tong University 2030 Initiative, the Shanghai Pilot Program for Basic Research and the Fundamental Research Funds for the Central Universities.  MM, DP, and MMW acknowledge support from the National Science Foundation under Grant Nos.~2329662 and~2611810.

The authors used the AI assistant ChatGPT Astra (OpenAI) in this work. This tool assisted with the scientific reasoning and derivations behind Theorem~\ref{thm:fock} and Propositions~\ref{prop:relent} and~\ref{prop:dpi}, including suggesting the second-quantization formulation, drafting the proofs in the Supplemental Material, identifying related literature, and drafting and revising the text. The authors directed the tools through iterative prompts that specified the questions to be addressed and the checks to be performed. The authors verified all derivations and proofs by hand, checked the cited references, and revised all AI-generated text, and they take full responsibility for the content of this Letter.
\end{acknowledgments}

\bibliography{ref-fock}

\clearpage
\onecolumngrid
\raggedbottom

\large

\setcounter{equation}{0}
\renewcommand{\theequation}{S\arabic{equation}}
\renewcommand{\theHequation}{S\arabic{equation}}
\begin{center}
\textbf{Supplemental Material for ``Semidefinite optimization as many-body thermodynamics:\\Boltzmann, Fermi--Dirac, and Bose--Einstein frameworks''}
\end{center}
\medskip

\phantomsection\label{app:A}
\section*{Appendix A: Fock space and one-body density matrices}

\textit{Fock space and mode operators.}---We use the standard formalism of second quantization~\cite{Bratteli1997}. As in the main text, $\Hil\simeq\mathbb{C}^{d}$ is the Hilbert space on which the SDP is defined, with orthonormal basis $\{|j\rangle\}_{j\in[d]}$; in the many-body picture, $\Hil$ is the single-particle Hilbert space, and each basis vector $|j\rangle$ is a mode. The Fock space $\Fock_{\pm}(\Hil)$ is spanned by the occupation-number states $|n_{1},\dots,n_{d}\rangle$, in which mode $j$ holds $n_{j}$ particles, with $n_{j}\in\{0,1\}$ for fermions ($\Fock_{-}$) and $n_{j}\in\{0,1,2,\dots\}$ for bosons ($\Fock_{+}$). The annihilation operator $a_{j}$ removes a particle from mode $j$, and its adjoint, the creation operator $a_{j}^{\dagger}$, adds one; for bosons,
\begin{equation}\label{eq:S-ladder}
a_{j}|\dots,n_{j},\dots\rangle=\sqrt{n_{j}}\,|\dots,n_{j}-1,\dots\rangle,\qquad a_{j}^{\dagger}|\dots,n_{j},\dots\rangle=\sqrt{n_{j}+1}\,|\dots,n_{j}+1,\dots\rangle,
\end{equation}
and, for fermions, the same holds up to the sign $(-1)^{n_{1}+\dots+n_{j-1}}$, except that $a_{j}^{\dagger}$ annihilates the states with $n_{j}=1$. In particular, $a_{j}^{\dagger}a_{j}|n_{1},\dots,n_{d}\rangle=n_{j}|n_{1},\dots,n_{d}\rangle$, so that $a_{j}^{\dagger}a_{j}$ counts the number of particles in mode $j$, and $\Nhat=\sum_{j}a_{j}^{\dagger}a_{j}$ is the particle-number operator. These operators satisfy the canonical anticommutation relations (fermions) and commutation relations (bosons),
\begin{equation}\label{eq:S-car-basis}
\{a_{j},a_{k}^{\dagger}\}=\delta_{jk}I,\ \ \{a_{j},a_{k}\}=0\ \ \text{(fermions)},\qquad [a_{j},a_{k}^{\dagger}]=\delta_{jk}I,\ \ [a_{j},a_{k}]=0\ \ \text{(bosons)}.
\end{equation}
For bosons, $\|a_{j}^{\dagger}|\dots,n_{j},\dots\rangle\|=\sqrt{n_{j}+1}$ grows without bound, so that $a_{j}$ and $a_{j}^{\dagger}$ are unbounded operators; we therefore consider states with finite mean particle number, $\omega(\Nhat)<\infty$, for which all the expectation values below are finite.

\textit{Expectation values, entropy, and relative entropy.}---A state $\omega$ on $\Fock_{\pm}(\Hil)$ is a density operator, $\omega\succeq0$ with $\Tr[\omega]=1$, and for an operator $X$ on $\Fock_{\pm}(\Hil)$ we write
\begin{equation}\label{eq:S-expect}
\omega(X)\coloneqq\Tr[X \omega]
\end{equation}
for the expectation value of $X$ in the state $\omega$. Thus $\omega(X)$ is a number, and $X\mapsto\omega(X)$ is a linear function of $X$ that is normalized, $\omega(I)=1$, and positive: if $X=B^{\dagger}B$ for some operator $B$, then, writing $\omega=\sum_{i}p_{i}|\psi_{i}\rangle\!\langle\psi_{i}|$ in its spectral decomposition, with $p_{i}\ge0$ and orthonormal $\psi_{i}$,
\begin{equation}\label{eq:S-positive}
\omega(B^{\dagger}B)=\sum_{i}p_{i}\,\langle\psi_{i}|B^{\dagger}B|\psi_{i}\rangle=\sum_{i}p_{i}\,\|B\psi_{i}\|^{2}\ \ge\ 0 .
\end{equation}
We also write $S(\omega)\coloneqq-\Tr[\omega\ln\omega]$ for the von Neumann entropy and $D(\omega\|\omega')\coloneqq\Tr[\omega(\ln\omega-\ln\omega')]$ for the Umegaki relative entropy of states on the Fock space; $D(\omega\|\omega')\ge0$, with equality if and only if $\omega=\omega'$~\cite{Watrous2018}. 

\textit{Modes.}---For every vector $\varphi\in\Hil$, not necessarily normalized, we define
\begin{equation}\label{eq:S-modes}
a(\varphi)\coloneqq\sum_{j}\langle\varphi|j\rangle\,a_{j},\qquad a(\varphi)^{\dagger}=\sum_{j}\langle j|\varphi\rangle\,a_{j}^{\dagger},
\end{equation}
so that $a(\varphi)$ is antilinear and $a(\varphi)^{\dagger}$ is linear in $\varphi$. When $\varphi$ is a unit vector, $a(\varphi)$ and $a(\varphi)^{\dagger}$ annihilate and create, respectively, a particle in the single-particle state $\varphi$; for $\varphi=|j\rangle$, they reduce to $a_{j}$ and $a_{j}^{\dagger}$. By linearity, the relations~\eqref{eq:S-car-basis} become, for all $\varphi,\psi\in\Hil$,
\begin{equation}\label{eq:S-car}
\{a(\varphi),a(\psi)^{\dagger}\}=\langle\varphi|\psi\rangle\,I\quad\text{(fermions)},\qquad [a(\varphi),a(\psi)^{\dagger}]=\langle\varphi|\psi\rangle\,I\quad\text{(bosons)}.
\end{equation}
For a unit vector $\varphi$, the operator
\begin{equation}\label{eq:S-number}
\hat n_{\varphi}\coloneqq a(\varphi)^{\dagger}a(\varphi)
\end{equation}
is the number operator of the mode $\varphi$: its eigenvalues count the particles in the single-particle state $\varphi$. If $A=\sum_{m}\alpha_{m}|\varphi_{m}\rangle\!\langle\varphi_{m}|$ is a spectral decomposition, inserting $\langle j|A|k\rangle=\sum_{m}\alpha_{m}\langle j|\varphi_{m}\rangle\langle\varphi_{m}|k\rangle$ into~\eqref{eq:dGamma} and using~\eqref{eq:S-modes} gives
\begin{equation}\label{eq:S-dGamma-diag}
\dG(A)=\sum_{m}\alpha_{m}\,a(\varphi_{m})^{\dagger}a(\varphi_{m})=\sum_{m}\alpha_{m}\,\hat n_{\varphi_{m}},
\end{equation}
that is, $\dG(A)$ adds up the value $\alpha_{m}$ of $A$ in each mode $\varphi_{m}$, weighted by the number of particles in that mode.

\textit{One-body density matrix.}---The one-body density matrix $\gamma_{\omega}$ of a state $\omega$ is the $d\times d$ matrix of expectation values $\langle j|\gamma_{\omega}|k\rangle\coloneqq\omega(a_{k}^{\dagger}a_{j})$. By linearity of $\omega(\cdot)$ and~\eqref{eq:S-modes}, for all $\varphi,\psi\in\Hil$,
\begin{equation}\label{eq:A1}
\langle\varphi|\gamma_{\omega}|\psi\rangle=\sum_{j,k}\langle\varphi|j\rangle\langle k|\psi\rangle\,\omega(a_{k}^{\dagger}a_{j})=\omega\big(a(\psi)^{\dagger}a(\varphi)\big),
\end{equation}
and in particular, for a unit vector $\varphi$,
\begin{equation}\label{eq:S-occupation}
\langle\varphi|\gamma_{\omega}|\varphi\rangle=\omega(\hat n_{\varphi})
\end{equation}
is the mean number of particles in the single-particle state $\varphi$.  Likewise, for every Hermitian $A$ on $\Hil$,
\begin{equation}\label{eq:S-1rdm}
\omega\big(\dG(A)\big)=\sum_{j,k}\langle j|A|k\rangle\,\omega(a_{j}^{\dagger}a_{k})=\sum_{j,k}\langle j|A|k\rangle\langle k|\gamma_{\omega}|j\rangle=\Tr[A\gamma_{\omega}],
\end{equation}
which is Eq.~\eqref{eq:1rdm} of the main text; for $A=I$, it gives $\Tr[\gamma_{\omega}]=\omega(\Nhat)$, the mean particle number.

\begin{lemma}\label{lem:constraints}
For every state $\omega$ on $\Fock_{\pm}(\Hil)$ (with $\omega(\Nhat)<\infty$ for bosons), $\gamma_{\omega}\succeq0$; for fermions, moreover, $\gamma_{\omega}\preceq I$.
\end{lemma}
\begin{proof}
Let $\varphi\in\Hil$ be a unit vector. The number operator $\hat n_{\varphi}=a(\varphi)^{\dagger}a(\varphi)$ has the form $B^{\dagger}B$ with $B=a(\varphi)$, so that~\eqref{eq:S-occupation} and~\eqref{eq:S-positive} give
\begin{equation}
\langle\varphi|\gamma_{\omega}|\varphi\rangle=\omega\big(a(\varphi)^{\dagger}a(\varphi)\big)=\sum_{i}p_{i}\,\|a(\varphi)\psi_{i}\|^{2}\ \ge\ 0,
\end{equation}
i.e., a mean number of particles is nonnegative. As $\varphi$ is arbitrary, $\gamma_{\omega}\succeq0$. For fermions, \eqref{eq:S-car} with $\psi=\varphi$ gives
\begin{equation}
a(\varphi)^{\dagger}a(\varphi)=I-a(\varphi)a(\varphi)^{\dagger},
\end{equation}
so that normalization, $\omega(I)=1$, and~\eqref{eq:S-positive} with $B=a(\varphi)^{\dagger}$ give
\begin{equation}
\langle\varphi|\gamma_{\omega}|\varphi\rangle=\omega(I)-\omega\big(a(\varphi)a(\varphi)^{\dagger}\big)=1-\omega\big(a(\varphi)a(\varphi)^{\dagger}\big)\ \le\ 1 .
\end{equation}
As $\varphi$ is arbitrary, $\gamma_{\omega}\preceq I$. Equivalently, \eqref{eq:S-car} implies $a(\varphi)^{2}=0$, hence $\hat n_{\varphi}^{2}=a(\varphi)^{\dagger}\big(I-a(\varphi)^{\dagger}a(\varphi)\big)a(\varphi)=\hat n_{\varphi}$: the number operator of a fermionic mode is a projector, and the mode holds at most one particle (Pauli exclusion). For bosons, \eqref{eq:S-car} gives instead $\hat n_{\varphi}=a(\varphi)a(\varphi)^{\dagger}-I$, which is not bounded from above, and $\omega(\hat n_{\varphi})$ can take any value in $[0,\infty)$.
\end{proof}

Conversely, every element of $\mathcal{C}_{\mathrm{FD}}$, resp.\ $\mathcal{C}_{\mathrm{BE}}$, is the one-body density matrix of some state~\cite{Coleman1963}, e.g., of the quasi-free state of Lemma~\ref{lem:quasifree} for interior points; this converse is not needed below.

\begin{lemma}\label{lem:quasifree}
Let $\Lambda$ be a Hermitian operator on $\Hil$ with spectral decomposition $\Lambda=\sum_{m}\lambda_{m}|\varphi_{m}\rangle\!\langle\varphi_{m}|$, where the $\lambda_{m}\in\mathbb{R}$ are its eigenvalues and $\{|\varphi_{m}\rangle\}_{m\in[d]}$ is an orthonormal basis of eigenvectors, with $\Lambda\succ0$ (i.e., $\lambda_{m}>0$ for all $m$) for bosons, and let $\omega_{\Lambda}\coloneqq e^{-\dG(\Lambda)}/Z(\Lambda)$ be the corresponding (gauge-invariant) quasi-free state, with $Z(\Lambda)\coloneqq\Tr[e^{-\dG(\Lambda)}]$. Then the following hold.
\begin{enumerate}
\item[(i)] Partition function: $Z(\Lambda)=\det(I+e^{-\Lambda})$ for fermions and $Z(\Lambda)=\det(I-e^{-\Lambda})^{-1}$ for bosons.
\item[(ii)] One-body density matrix: $\gamma_{\omega_{\Lambda}}=(e^{\Lambda}+I)^{-1}$ for fermions and $\gamma_{\omega_{\Lambda}}=(e^{\Lambda}-I)^{-1}$ for bosons.
\item[(iii)] Entropy: $S(\omega_{\Lambda})=S_{\mathrm{FD}}(\gamma_{\omega_{\Lambda}})$ for fermions and $S(\omega_{\Lambda})=S_{\mathrm{BE}}(\gamma_{\omega_{\Lambda}})$ for bosons.
\item[(iv)] Entropy deficit: for every state $\omega$ with the same one-body density matrix as $\omega_{\Lambda}$, i.e., with $\gamma_{\omega}=\gamma_{\omega_{\Lambda}}$, it holds that $S(\omega)=S(\omega_{\Lambda})-D(\omega\|\omega_{\Lambda})$, so that $S(\omega)\le S(\omega_{\Lambda})$.
\end{enumerate}
\end{lemma}

\begin{proof}
\textit{Factorization over modes.} Let $b_{m}\coloneqq a(\varphi_{m})$ and $\hat n_{m}\coloneqq b_{m}^{\dagger}b_{m}$. By~\eqref{eq:S-dGamma-diag},
\begin{equation}
\dG(\Lambda)=\sum_{m}\lambda_{m}\,\hat n_{m}.
\end{equation}
The mode-number operators $\hat n_{m}$ commute, and their joint eigenvectors are the occupation-number states $|n_{1},\dots,n_{d}\rangle$ of the modes $\varphi_{1},\dots,\varphi_{d}$, with
\begin{equation}
\hat n_{m}|n_{1},\dots,n_{d}\rangle=n_{m}|n_{1},\dots,n_{d}\rangle,\qquad n_{m}\in\{0,1\}\ \text{(fermions)},\quad n_{m}\in\{0,1,2,\dots\}\ \text{(bosons)}.
\end{equation}
Hence
\begin{equation}
e^{-\dG(\Lambda)}|n_{1},\dots,n_{d}\rangle=\prod_{m}e^{-\lambda_{m}n_{m}}\,|n_{1},\dots,n_{d}\rangle,
\end{equation}
so that $\omega_{\Lambda}$ is diagonal in the occupation-number basis, with eigenvalues
\begin{equation}\label{eq:S-product}
\langle n_{1},\dots,n_{d}|\omega_{\Lambda}|n_{1},\dots,n_{d}\rangle=\prod_{m}p_{m}(n_{m}),\qquad p_{m}(n)\coloneqq\frac{e^{-\lambda_{m}n}}{z(\lambda_{m})},
\end{equation}
where the single-mode partition function $z(\lambda)\coloneqq\sum_{n}e^{-\lambda n}$ is a sum over the allowed occupation numbers $n$ of one mode. For fermions, $n\in\{0,1\}$, so that
\begin{equation}
z(\lambda)=e^{0}+e^{-\lambda}=1+e^{-\lambda}.
\end{equation}
For bosons, $n\in\{0,1,2,\dots\}$, and $z(\lambda)$ is a geometric series with ratio $x\coloneqq e^{-\lambda}$, which satisfies $0<x<1$ because $\lambda>0$ (recall that $\Lambda\succ0$ for bosons); since $(1-x)\sum_{n=0}^{N}x^{n}=1-x^{N+1}\to1$ as $N\to\infty$,
\begin{equation}
z(\lambda)=\sum_{n=0}^{\infty}x^{n}=\frac{1}{1-x}=\frac{1}{1-e^{-\lambda}}.
\end{equation}
Each mode is thus occupied independently, according to the distribution $p_{m}$.

\textit{(i)} Summing the unnormalized eigenvalues,
\begin{equation}
Z(\Lambda)=\sum_{n_{1},\dots,n_{d}}\prod_{m}e^{-\lambda_{m}n_{m}}=\prod_{m}z(\lambda_{m})=\begin{cases}\prod_{m}(1+e^{-\lambda_{m}})=\det(I+e^{-\Lambda}), & \text{fermions},\\ \prod_{m}(1-e^{-\lambda_{m}})^{-1}=\det(I-e^{-\Lambda})^{-1}, & \text{bosons}.\end{cases}
\end{equation}

\textit{(ii)} The mean occupation of the mode $\varphi_{m}$ is
\begin{equation}
\bar n_{m}\coloneqq\omega_{\Lambda}(\hat n_{m})=\sum_{n}n\,p_{m}(n)=\begin{cases}\dfrac{e^{-\lambda_{m}}}{1+e^{-\lambda_{m}}}=\dfrac{1}{e^{\lambda_{m}}+1}, & \text{fermions},\\[2ex] \dfrac{e^{-\lambda_{m}}}{1-e^{-\lambda_{m}}}=\dfrac{1}{e^{\lambda_{m}}-1}, & \text{bosons},\end{cases}
\end{equation}
where, for bosons, we used $\sum_{n}n\,x^{n}=x\,\frac{d}{dx}\sum_{n}x^{n}=x/(1-x)^{2}$ with $x=e^{-\lambda_{m}}$, so that $\bar n_{m}=(1-x)\,x/(1-x)^{2}=x/(1-x)$. Moreover, for $m\neq l$, $b_{m}^{\dagger}b_{l}$ moves a particle from $\varphi_{l}$ to $\varphi_{m}$ and thus maps every occupation-number state to an orthogonal one, so that $\omega_{\Lambda}(b_{m}^{\dagger}b_{l})=0$, because $\omega_{\Lambda}$ is diagonal in that basis. By~\eqref{eq:A1},
\begin{equation}
\langle\varphi_{l}|\gamma_{\omega_{\Lambda}}|\varphi_{m}\rangle=\omega_{\Lambda}(b_{m}^{\dagger}b_{l})=\delta_{lm}\,\bar n_{m},\qquad\text{i.e.,}\qquad\gamma_{\omega_{\Lambda}}=\sum_{m}\bar n_{m}|\varphi_{m}\rangle\!\langle\varphi_{m}|=\begin{cases}(e^{\Lambda}+I)^{-1}, & \text{fermions},\\ (e^{\Lambda}-I)^{-1}, & \text{bosons}.\end{cases}
\end{equation}

\textit{(iii)} Because the eigenvalues~\eqref{eq:S-product} of $\omega_{\Lambda}$ form a product distribution, its entropy is the sum of the Shannon entropies of the single-mode distributions:
\begin{equation}
S(\omega_{\Lambda})=-\sum_{n_{1},\dots,n_{d}}\prod_{m}p_{m}(n_{m})\,\ln\prod_{m}p_{m}(n_{m})=\sum_{m}h(p_{m}),\qquad h(p)\coloneqq-\sum_{n}p(n)\ln p(n).
\end{equation}
A fermionic mode is occupied with probability $\bar n_{m}$, so that $p_{m}=(1-\bar n_{m},\bar n_{m})$, and a bosonic mode has the geometric distribution $p_{m}(n)=(1-q_{m})\,q_{m}^{n}$ with $q_{m}=e^{-\lambda_{m}}=\bar n_{m}/(\bar n_{m}+1)$, so that
\begin{equation}
h(p_{m})=\begin{cases}-\bar n_{m}\ln\bar n_{m}-(1-\bar n_{m})\ln(1-\bar n_{m}), & \text{fermions},\\ -\ln(1-q_{m})-\bar n_{m}\ln q_{m}=(\bar n_{m}+1)\ln(\bar n_{m}+1)-\bar n_{m}\ln\bar n_{m}, & \text{bosons}.\end{cases}
\end{equation}
Since the $\bar n_{m}$ are the eigenvalues of $\gamma_{\omega_{\Lambda}}$ by (ii), summing over $m$ and comparing with the entropies of Table~\ref{tab:three-frameworks} gives $S(\omega_{\Lambda})=S_{\mathrm{FD}}(\gamma_{\omega_{\Lambda}})$ for fermions and $S(\omega_{\Lambda})=S_{\mathrm{BE}}(\gamma_{\omega_{\Lambda}})$ for bosons (see also Refs.~\cite{BachLiebSolovej1994,Dierckx2008}).

\textit{(iv)} Since $\ln\omega_{\Lambda}=-\dG(\Lambda)-\ln Z(\Lambda)\,I$, Eq.~\eqref{eq:S-1rdm} gives, for every state $\omega$,
\begin{equation}\label{eq:A2}
-\Tr[\omega\ln\omega_{\Lambda}]=\omega\big(\dG(\Lambda)\big)+\ln Z(\Lambda)=\Tr[\Lambda\,\gamma_{\omega}]+\ln Z(\Lambda),
\end{equation}
a quantity that depends on $\omega$ only through $\gamma_{\omega}$. Hence, if $\gamma_{\omega}=\gamma_{\omega_{\Lambda}}$, then $-\Tr[\omega\ln\omega_{\Lambda}]=-\Tr[\omega_{\Lambda}\ln\omega_{\Lambda}]=S(\omega_{\Lambda})$, and
\begin{equation}
D(\omega\|\omega_{\Lambda})=\Tr[\omega\ln\omega]-\Tr[\omega\ln\omega_{\Lambda}]=-S(\omega)+S(\omega_{\Lambda}),
\end{equation}
which is (iv).
\end{proof}

\textit{Remark.}---Item~(iv) concerns all the states that share the one-body density matrix of $\omega_{\Lambda}$, and there are many of them: $\gamma_{\omega}$ fixes only the $d^{2}$ expectation values $\omega(a_{k}^{\dagger}a_{j})$ of the one-body operators $a_{k}^{\dagger}a_{j}$, whereas a state on the Fock space depends on many more parameters, which encode, e.g., the correlations between different modes. For example, for $d=2$ fermionic modes, the Fock space is spanned by $|00\rangle$, $|10\rangle$, $|01\rangle$, and $|11\rangle$, and the three states
\begin{equation}
\omega_{\Lambda=0}=\frac{1}{4}\,I,\qquad\frac{1}{2}\big(|10\rangle\!\langle10|+|01\rangle\!\langle01|\big),\qquad\frac{1}{2}\big(|00\rangle+|11\rangle\big)\big(\langle00|+\langle11|\big)
\end{equation}
all have the same one-body density matrix $\gamma=\frac{1}{2}I$, but different entropies, namely $2\ln2$, $\ln2$, and $0$, respectively. The first is the quasi-free state $\omega_{\Lambda}$ with $\Lambda=0$, the second describes exactly one particle, found in either mode with equal probability, and the third is a pure superposition of the empty and the doubly occupied configurations (a quasi-free state only in the broader sense that allows pairing, $\omega(a_{1}a_{2})\neq0$, and not gauge-invariant). Item~(iv) states that, among all these states, the gauge-invariant quasi-free one has the largest entropy, here $S_{\mathrm{FD}}(\frac{1}{2}I)=2\ln2$, and that the entropy of every other state $\omega$ falls short of it by $D(\omega\|\omega_{\Lambda})$.

\textit{Maximum-entropy property.}---Given $\gamma$ in the interior of $\mathcal{C}_{\stat}$, inverting Lemma~\ref{lem:quasifree}(ii) gives
\begin{equation}
\gamma_{\omega_{\Lambda}}=\gamma\quad\Longleftrightarrow\quad e^{\Lambda}=\gamma^{-1}\mp I\quad\Longleftrightarrow\quad \Lambda=\Lambda_{\gamma},
\end{equation}
with $\Lambda_{\gamma}$ as in Eq.~\eqref{eq:quasifree} of the main text (upper sign for fermions, lower sign for bosons, in which case $\Lambda_{\gamma}\succ0$). We write $\omega_{\gamma}\coloneqq\omega_{\Lambda_{\gamma}}$ for this quasi-free state; thus a subscript $\gamma$, $\sigma$, or $X$ labels a quasi-free state by its one-body density matrix, whereas a subscript $\Lambda$ or $K/T$ labels it by its one-body Hamiltonian. For $\gamma$ on the boundary of $\mathcal{C}_{\stat}$, $\omega_{\gamma}$ denotes the state in which the modes in $\ker\gamma$ are empty, those in the eigenvalue-one eigenspace of $\gamma$ (for fermions) are filled, and the remaining modes are in the quasi-free state of the restriction of $\gamma$. For every state $\omega$ with $\gamma_{\omega}=\gamma$, Lemma~\ref{lem:quasifree}(iv) and~(iii) then give
\begin{equation}\label{eq:maxent}
S(\omega)=S(\omega_{\gamma})-D(\omega\|\omega_{\gamma})=S_{\stat}(\gamma)-D(\omega\|\omega_{\gamma})\ \le\ S_{\stat}(\gamma),
\end{equation}
with equality if and only if $\omega=\omega_{\gamma}$, by faithfulness of the relative entropy; this is the maximum-entropy property of quasi-free states~\cite{BachLiebSolovej1994,WolfGiedkeCirac2006}.

The inequality in~\eqref{eq:maxent} also holds if $\gamma_{\omega}$ lies on the boundary of $\mathcal{C}_{\stat}$. In that case, there is a unit vector $\varphi$ with $\gamma_{\omega}\varphi=0$, or, for fermions, $\gamma_{\omega}\varphi=\varphi$. In the first case, \eqref{eq:S-occupation} and~\eqref{eq:S-positive} give $0=\omega(\hat n_{\varphi})=\sum_{i}p_{i}\|a(\varphi)\psi_{i}\|^{2}$, so that $a(\varphi)\psi_{i}=0$ for every eigenvector $\psi_{i}$ of $\omega$ with $p_{i}>0$: the mode $\varphi$ is empty. In the second case, the same argument applied to $\omega(a(\varphi)a(\varphi)^{\dagger})=1-\omega(\hat n_{\varphi})=0$ gives $a(\varphi)^{\dagger}\psi_{i}=0$: the mode $\varphi$ is filled. Identifying $\Fock_{\pm}(\Hil)\simeq\Fock_{\pm}(\mathbb{C}\varphi)\otimes\Fock_{\pm}(\varphi^{\perp})$, it follows that
\begin{equation}
\omega=|n\rangle\!\langle n|\otimes\omega',\qquad \gamma_{\omega}=n\,|\varphi\rangle\!\langle\varphi|\oplus\gamma_{\omega'},\qquad n\in\{0,1\}\ \ (n=0\ \text{for bosons}),
\end{equation}
with $\omega'$ a state of the remaining modes. Hence $S(\omega)=S(\omega')$, while $S_{\stat}(\gamma_{\omega})=S_{\stat}(\gamma_{\omega'})$, because the single-mode entropy vanishes for an empty mode and, for fermions, for a filled one. Iterating over an orthonormal basis of $\ker\gamma_{\omega}$ (and, for fermions, of $\ker(I-\gamma_{\omega})$) reduces the inequality in~\eqref{eq:maxent} to the interior case.

\phantomsection\label{app:B}
\section*{Appendix B: Proof of Theorem~\ref{thm:fock}}

\textit{Boltzmann case.}---For $\stat=\mathrm{B}$, we have $\Fock_{\mathrm{B}}=\Hil$, $\dG_{\mathrm{B}}(A)=A$, and $\gamma_{\omega}=\omega$, so that~\eqref{eq:fock-free-energy} is~\eqref{eq:freeen} itself, whose unique minimizer is the Gibbs state $e^{-K_{\mu^{\star}}/T}/Z_{\mathrm{B}}(\mu^{\star})$ and whose Lagrangian dual function is~\eqref{eq:dual-Z}~\cite{liu2025sdp}. 

Now let $\stat\in\{\mathrm{FD},\mathrm{BE}\}$ and, for a Hermitian operator $K$ on $\Hil$ (with $K\succ0$ for bosons), let $\omega_{K/T}$ denote the state of Lemma~\ref{lem:quasifree} with $\Lambda=K/T$.

\textit{Step 1: free energy without constraints.}---We show that
\begin{equation}\label{eq:B1}
\min_{X\in\mathcal{C}_{\stat}}\big\{\Tr[KX]-T S_{\stat}(X)\big\}=\min_{\omega}\big\{\omega\big(\dG(K)\big)-T S(\omega)\big\}=-T\ln Z(K/T),
\end{equation}
with the minima attained at $X=\gamma_{\omega_{K/T}}$ and $\omega=\omega_{K/T}$, respectively. The second equality is the Gibbs variational principle: Eq.~\eqref{eq:A2} with $\Lambda=K/T$ gives, for every state $\omega$,
\begin{equation}
T\,D(\omega\|\omega_{K/T})=T\,\Tr[\omega\ln\omega]-T\,\Tr[\omega\ln\omega_{K/T}]=-T S(\omega)+\omega\big(\dG(K)\big)+T\ln Z(K/T),
\end{equation}
that is,
\begin{equation}\label{eq:B-gibbs}
\omega\big(\dG(K)\big)-T S(\omega)=-T\ln Z(K/T)+T\,D(\omega\|\omega_{K/T})\ \ge\ -T\ln Z(K/T),
\end{equation}
with equality if and only if $\omega=\omega_{K/T}$. For the first equality, let $\omega$ be any state. By Lemma~\ref{lem:constraints}, $\gamma_{\omega}\in\mathcal{C}_{\stat}$; by~\eqref{eq:S-1rdm}, $\omega(\dG(K))=\Tr[K\gamma_{\omega}]$; and by~\eqref{eq:maxent}, $S(\omega)\le S_{\stat}(\gamma_{\omega})$. Hence
\begin{equation}
\omega\big(\dG(K)\big)-T S(\omega)\ \ge\ \Tr[K\gamma_{\omega}]-T S_{\stat}(\gamma_{\omega})\ \ge\ \min_{X\in\mathcal{C}_{\stat}}\big\{\Tr[KX]-T S_{\stat}(X)\big\}.
\end{equation}
Conversely, for every $X$ in the interior of $\mathcal{C}_{\stat}$, the quasi-free state $\omega_{X}$ satisfies $\gamma_{\omega_{X}}=X$ and $S(\omega_{X})=S_{\stat}(X)$ by Lemma~\ref{lem:quasifree}(ii)--(iii), so that
\begin{equation}
\Tr[KX]-T S_{\stat}(X)=\omega_{X}\big(\dG(K)\big)-T S(\omega_{X})\ \ge\ \min_{\omega}\big\{\omega\big(\dG(K)\big)-T S(\omega)\big\},
\end{equation}
and the same holds on the boundary of $\mathcal{C}_{\stat}$, whose points are limits of interior points along which the objective is continuous. The two minima in~\eqref{eq:B1} are therefore equal, and their common value $-T\ln Z(K/T)$ is attained at $\omega=\omega_{K/T}$ and at $X=\gamma_{\omega_{K/T}}$, because $\omega_{K/T}$ is itself quasi-free.

\textit{Step 2: free energy with constraints.}---Let $\omega$ be feasible for the right-hand side of~\eqref{eq:fock-free-energy}, i.e., $\omega(\dG(Q_{i}))=q_{i}$ for all $i\in[c]$. By Lemma~\ref{lem:constraints} and~\eqref{eq:S-1rdm},
\begin{equation}
\gamma_{\omega}\in\mathcal{C}_{\stat},\qquad \Tr[Q_{i}\gamma_{\omega}]=\omega\big(\dG(Q_{i})\big)=q_{i}\quad\forall i\in[c],
\end{equation}
so that $\gamma_{\omega}$ is feasible for~\eqref{eq:freeen}, and, by~\eqref{eq:S-1rdm} and~\eqref{eq:maxent},
\begin{equation}\label{eq:B2}
\omega\big(\dG(H)\big)-T S(\omega)\ \ge\ \Tr[H\gamma_{\omega}]-T S_{\stat}(\gamma_{\omega})\ \ge\ F_{T}^{\stat}(\mathcal{Q},q).
\end{equation}
Conversely, let $X^{\star}\coloneqq X_{T}^{\stat}(\mu^{\star})$ be the optimum of~\eqref{eq:freeen}~\cite{liu2026sdp_fermi,minervini2026sdp_bose}, which lies in the interior of $\mathcal{C}_{\stat}$, and let $\omega^{\star}\coloneqq\omega_{K_{\mu^{\star}}/T}=\omega_{T}^{\stat}(\mu^{\star})$. By Lemma~\ref{lem:quasifree}(ii) with $\Lambda=K_{\mu^{\star}}/T$ (upper sign for fermions, lower sign for bosons),
\begin{equation}
\gamma_{\omega^{\star}}=\big(e^{K_{\mu^{\star}}/T}\pm I\big)^{-1}=X^{\star},
\end{equation}
so that $\omega^{\star}(\dG(Q_{i}))=\Tr[Q_{i}X^{\star}]=q_{i}$ by~\eqref{eq:S-1rdm}, i.e., $\omega^{\star}$ is feasible, and, by Lemma~\ref{lem:quasifree}(iii),
\begin{equation}
\omega^{\star}\big(\dG(H)\big)-T S(\omega^{\star})=\Tr[HX^{\star}]-T S_{\stat}(X^{\star})=F_{T}^{\stat}(\mathcal{Q},q).
\end{equation}
Together with~\eqref{eq:B2}, this proves~\eqref{eq:fock-free-energy} and shows that $\omega^{\star}$ is a minimizer. If $\omega$ is any minimizer, both inequalities in~\eqref{eq:B2} are equalities. The second makes $\gamma_{\omega}$ an optimum of~\eqref{eq:freeen}, so that $\gamma_{\omega}=X^{\star}$ by strict concavity of $S_{\stat}$~\cite{liu2026sdp_fermi,minervini2026sdp_bose}; the first gives $D(\omega\|\omega_{X^{\star}})=0$ by~\eqref{eq:maxent}, so that $\omega=\omega_{X^{\star}}=\omega^{\star}$.

\textit{Step 3: dual function.}---The Lagrangian of~\eqref{eq:freeen} is
\begin{equation}
\Tr[HX]-T S_{\stat}(X)+\sum_{i}\mu_{i}\big(q_{i}-\Tr[Q_{i}X]\big)=\mu\cdot q+\Tr[\Km X]-T S_{\stat}(X),
\end{equation}
so that its dual function is~\cite{liu2026sdp_fermi,minervini2026sdp_bose}
\begin{equation}
f_{T}^{\stat}(\mu)=\mu\cdot q+\min_{X\in\mathcal{C}_{\stat}}\big\{\Tr[\Km X]-T S_{\stat}(X)\big\}=\mu\cdot q-T\ln Z(\Km/T)=\mu\cdot q-T\ln Z_{\stat}(\mu),
\end{equation}
by~\eqref{eq:B1} with $K=\Km$, which is~\eqref{eq:dual-Z}. Finally, by linearity of $\dG$,
\begin{equation}
\dG(\Km)=\dG(H)-\sum_{i}\mu_{i}\,\dG(Q_{i}),
\end{equation}
so that $Z_{\stat}(\mu)$ is the grand partition function of the many-body system with Hamiltonian $\dG(H)$ and charges $\dG(Q_{i})$, and~\eqref{eq:dual-Z} is the dual function of Ref.~\cite{liu2025sdp} for that system.

\textit{Step 4: gradient and Hessian.}---The concavity of $f_{T}^{\stat}$, the gradient~\eqref{eq:grad}, and the Hessian~\eqref{eq:hessian} then follow from Ref.~\cite{liu2025sdp} applied to this many-body system; for completeness, we derive them here, for every $\stat\in\{\mathrm{B},\mathrm{FD},\mathrm{BE}\}$. Write $\hat Q_{i}\coloneqq\dG_{\stat}(Q_{i})$, $\omega_{\mu}\coloneqq\omega_{T}^{\stat}(\mu)$, and
\begin{equation}
A(\mu)\coloneqq-\frac{1}{T}\,\dG_{\stat}(\Km)=-\frac{1}{T}\Big(\dG_{\stat}(H)-\sum_{i}\mu_{i}\hat Q_{i}\Big),
\end{equation}
so that
\begin{equation}
Z_{\stat}(\mu)=\Tr\big[e^{A(\mu)}\big],\qquad\omega_{\mu}=\frac{e^{A(\mu)}}{Z_{\stat}(\mu)},\qquad\frac{\partial A}{\partial\mu_{i}}=\frac{\hat Q_{i}}{T}.
\end{equation}
(For bosons, all the traces below are finite because $\Km\succ0$.) The derivative of the exponential of an operator-valued function is given by Duhamel's formula,
\begin{equation}\label{eq:S-duhamel}
\frac{\partial}{\partial\mu_{j}}e^{A(\mu)}=\int_{0}^{1}ds\,e^{sA(\mu)}\,\frac{\partial A}{\partial\mu_{j}}\,e^{(1-s)A(\mu)}.
\end{equation}
Taking the trace and using its cyclicity gives
\begin{equation}
\frac{\partial}{\partial\mu_{i}}\ln Z_{\stat}(\mu)=\frac{1}{Z_{\stat}(\mu)}\Tr\!\Big[e^{A(\mu)}\,\frac{\hat Q_{i}}{T}\Big]=\frac{\omega_{\mu}(\hat Q_{i})}{T},
\end{equation}
so that, by~\eqref{eq:dual-Z}, \eqref{eq:S-1rdm}, and $\gamma_{\omega_{\mu}}=X_{T}^{\stat}(\mu)$ (Lemma~\ref{lem:quasifree}(ii) for $\stat\in\{\mathrm{FD},\mathrm{BE}\}$, and $\gamma_{\omega}=\omega$ for $\stat=\mathrm{B}$),
\begin{equation}
\frac{\partial f_{T}^{\stat}}{\partial\mu_{i}}=q_{i}-T\,\frac{\partial}{\partial\mu_{i}}\ln Z_{\stat}(\mu)=q_{i}-\omega_{\mu}(\hat Q_{i})=q_{i}-\Tr\big[Q_{i}X_{T}^{\stat}(\mu)\big],
\end{equation}
which is~\eqref{eq:grad}. Differentiating $\omega_{\mu}(\hat Q_{i})=\Tr[\hat Q_{i}e^{A(\mu)}]/Z_{\stat}(\mu)$ once more, using~\eqref{eq:S-duhamel} and $e^{sA(\mu)}=Z_{\stat}(\mu)^{s}\,\omega_{\mu}^{s}$, gives
\begin{align}
\frac{\partial}{\partial\mu_{j}}\,\omega_{\mu}(\hat Q_{i})&=\frac{1}{Z_{\stat}(\mu)}\int_{0}^{1}ds\,\Tr\!\Big[\hat Q_{i}\,e^{sA(\mu)}\,\frac{\hat Q_{j}}{T}\,e^{(1-s)A(\mu)}\Big]-\omega_{\mu}(\hat Q_{i})\,\frac{\partial}{\partial\mu_{j}}\ln Z_{\stat}(\mu)\notag\\
&=\frac{1}{T}\Big(\int_{0}^{1}ds\,\Tr\big[\omega_{\mu}^{s}\,\hat Q_{j}\,\omega_{\mu}^{1-s}\,\hat Q_{i}\big]-\omega_{\mu}(\hat Q_{i})\,\omega_{\mu}(\hat Q_{j})\Big)\notag\\
&=\frac{1}{T}\Cov^{\mathrm{KM}}_{\omega_{\mu}}(\hat Q_{j},\hat Q_{i}),
\end{align}
and therefore $\partial^{2}f_{T}^{\stat}/\partial\mu_{i}\partial\mu_{j}=-\partial\omega_{\mu}(\hat Q_{i})/\partial\mu_{j}$ is~\eqref{eq:hessian}; the Kubo--Mori covariance is symmetric in its two arguments, as the substitution $s\to1-s$ and the cyclicity of the trace show. Finally, for real coefficients $(c_{i})_{i}$, the Hermitian operator $C\coloneqq\sum_{i}c_{i}\hat Q_{i}-\omega_{\mu}\big(\sum_{i}c_{i}\hat Q_{i}\big)I$ satisfies
\begin{equation}
\sum_{i,j}c_{i}c_{j}\Cov^{\mathrm{KM}}_{\omega_{\mu}}(\hat Q_{i},\hat Q_{j})=\int_{0}^{1}ds\,\Tr\big[\omega_{\mu}^{s}\,C\,\omega_{\mu}^{1-s}\,C\big]=\int_{0}^{1}ds\,\big\|\omega_{\mu}^{s/2}\,C\,\omega_{\mu}^{(1-s)/2}\big\|_{2}^{2}\ \ge\ 0,
\end{equation}
where $\|Y\|_{2}^{2}\coloneqq\Tr[Y^{\dagger}Y]$. Hence the Hessian~\eqref{eq:hessian} is negative semidefinite, and $f_{T}^{\stat}$ is concave. $\square$

\phantomsection\label{app:C}
\section*{Appendix C: Proofs of Propositions~\ref{prop:relent} and~\ref{prop:dpi}, and channels on $\Hil$}

Throughout, $D(P\|R)\coloneqq\Tr[P(\ln P-\ln R)]$ also for unnormalized $P,R\succeq0$, as in Table~\ref{tab:three-frameworks}, with $0\ln0\coloneqq0$ and $D(P\|R)\coloneqq+\infty$ if $\operatorname{supp}P\not\subseteq\operatorname{supp}R$.

\textit{Proof of Proposition~\ref{prop:relent}.}---Let $\gamma,\sigma$ be interior points of $\mathcal{C}_{\stat}$, and let $\omega$ be a state with $\gamma_{\omega}=\gamma$. We first show that the difference $D(\omega\|\omega_{\sigma})-D(\omega\|\omega_{\gamma})$ depends on $\omega$ only through its one-body density matrix $\gamma$, so that it can be evaluated at the quasi-free state $\omega_{\gamma}$, and we then compute $D(\omega_{\gamma}\|\omega_{\sigma})$ explicitly.

\textit{Step 1: reduction to quasi-free states.} The logarithm of a quasi-free state is a one-body operator plus a constant, $\ln\omega_{\Lambda}=-\dG(\Lambda)-\ln Z(\Lambda)\,I$, so that its expectation value in any state depends on that state only through its one-body density matrix; this is Eq.~\eqref{eq:A2}, $\Tr[\omega\ln\omega_{\Lambda}]=-\Tr[\Lambda\,\gamma_{\omega}]-\ln Z(\Lambda)$. The states $\omega$ and $\omega_{\gamma}$ have the same one-body density matrix, $\gamma_{\omega}=\gamma=\gamma_{\omega_{\gamma}}$, the first by assumption and the second by the definition of $\omega_{\gamma}$. We apply~\eqref{eq:A2} twice: once with $\Lambda=\Lambda_{\sigma}$, for which $\omega_{\Lambda}=\omega_{\sigma}$, and once with $\Lambda=\Lambda_{\gamma}$, for which $\omega_{\Lambda}=\omega_{\gamma}$. Each time, the right-hand side of~\eqref{eq:A2} takes the same value for $\omega$ and for $\omega_{\gamma}$, so that we can replace $\omega$ by $\omega_{\gamma}$:
\begin{align}
\Tr[\omega\ln\omega_{\sigma}]&=-\Tr[\Lambda_{\sigma}\gamma]-\ln Z(\Lambda_{\sigma})=\Tr[\omega_{\gamma}\ln\omega_{\sigma}],\\
\Tr[\omega\ln\omega_{\gamma}]&=-\Tr[\Lambda_{\gamma}\gamma]-\ln Z(\Lambda_{\gamma})=\Tr[\omega_{\gamma}\ln\omega_{\gamma}].
\end{align}
In the difference of the relative entropies $D(\omega\|\omega_{\sigma})=\Tr[\omega\ln\omega]-\Tr[\omega\ln\omega_{\sigma}]$ and $D(\omega\|\omega_{\gamma})=\Tr[\omega\ln\omega]-\Tr[\omega\ln\omega_{\gamma}]$, the terms $\Tr[\omega\ln\omega]$ cancel, and the two identities above give
\begin{align}
D(\omega\|\omega_{\sigma})-D(\omega\|\omega_{\gamma})&=\Tr[\omega\ln\omega_{\gamma}]-\Tr[\omega\ln\omega_{\sigma}]\notag\\
&=\Tr[\omega_{\gamma}\ln\omega_{\gamma}]-\Tr[\omega_{\gamma}\ln\omega_{\sigma}]\notag\\
&=D(\omega_{\gamma}\|\omega_{\sigma}).\label{eq:C0}
\end{align}
\textit{Step 2: explicit form of $D(\omega_{\gamma}\|\omega_{\sigma})$.} It remains to show that $D(\omega_{\gamma}\|\omega_{\sigma})=D_{\stat}(\gamma\|\sigma)$. By Lemma~\ref{lem:quasifree}(iii) and~\eqref{eq:A2} with $\Lambda=\Lambda_{\sigma}$,
\begin{equation}\label{eq:C1}
D(\omega_{\gamma}\|\omega_{\sigma})=-S(\omega_{\gamma})-\Tr[\omega_{\gamma}\ln\omega_{\sigma}]=-S_{\stat}(\gamma)+\Tr[\Lambda_{\sigma}\gamma]+\ln Z(\Lambda_{\sigma}).
\end{equation}
For fermions, $\Lambda_{\sigma}=\ln(\sigma^{-1}-I)=\ln(I-\sigma)-\ln\sigma$, so that $e^{-\Lambda_{\sigma}}=\sigma(I-\sigma)^{-1}$ and $I+e^{-\Lambda_{\sigma}}=(I-\sigma)^{-1}$; Lemma~\ref{lem:quasifree}(i) then gives $\ln Z(\Lambda_{\sigma})=\ln\det(I+e^{-\Lambda_{\sigma}})=-\Tr[\ln(I-\sigma)]$. Inserting this and $S_{\mathrm{FD}}(\gamma)=-\Tr[\gamma\ln\gamma+(I-\gamma)\ln(I-\gamma)]$ into~\eqref{eq:C1},
\begin{align}
D(\omega_{\gamma}\|\omega_{\sigma})&=\Tr[\gamma\ln\gamma+(I-\gamma)\ln(I-\gamma)]+\Tr[\gamma\ln(I-\sigma)-\gamma\ln\sigma]-\Tr[\ln(I-\sigma)]\notag\\
&=\Tr[\gamma\ln\gamma-\gamma\ln\sigma]+\Tr[(I-\gamma)\ln(I-\gamma)-(I-\gamma)\ln(I-\sigma)]\notag\\
&=D(\gamma\|\sigma)+D(I-\gamma\|I-\sigma)\notag\\
& =D_{\mathrm{FD}}(\gamma\|\sigma).
\end{align}
For bosons, $\Lambda_{\sigma}=\ln(\sigma^{-1}+I)=\ln(\sigma+I)-\ln\sigma$, so that $e^{-\Lambda_{\sigma}}=\sigma(\sigma+I)^{-1}$ and $I-e^{-\Lambda_{\sigma}}=(\sigma+I)^{-1}$; Lemma~\ref{lem:quasifree}(i) then gives $\ln Z(\Lambda_{\sigma})=-\ln\det(I-e^{-\Lambda_{\sigma}})=\Tr[\ln(\sigma+I)]$. Inserting this and $S_{\mathrm{BE}}(\gamma)=\Tr[(\gamma+I)\ln(\gamma+I)-\gamma\ln\gamma]$ into~\eqref{eq:C1},
\begin{align}
D(\omega_{\gamma}\|\omega_{\sigma})&=\Tr[\gamma\ln\gamma-(\gamma+I)\ln(\gamma+I)]+\Tr[\gamma\ln(\sigma+I)-\gamma\ln\sigma]+\Tr[\ln(\sigma+I)]\notag\\
&=\Tr[\gamma\ln\gamma-\gamma\ln\sigma]-\Tr[(\gamma+I)\ln(\gamma+I)-(\gamma+I)\ln(\sigma+I)]\notag\\
&=D(\gamma\|\sigma)-D(\gamma+I\|\sigma+I)\notag\\
&=D_{\mathrm{BE}}(\gamma\|\sigma).
\end{align}
Equivalently, in Bregman form: setting $\gamma=\sigma$ in~\eqref{eq:C1} gives $\ln Z(\Lambda_{\sigma})=S_{\stat}(\sigma)-\Tr[\Lambda_{\sigma}\sigma]$, and $\Lambda_{\sigma}=\nabla S_{\stat}(\sigma)$, as the gradients $\nabla S_{\mathrm{FD}}(\sigma)=\ln(I-\sigma)-\ln\sigma$ and $\nabla S_{\mathrm{BE}}(\sigma)=\ln(\sigma+I)-\ln\sigma$ show~\cite{liu2026sdp_fermi,minervini2026sdp_bose}, so that
\begin{equation}
D(\omega_{\gamma}\|\omega_{\sigma})=-S_{\stat}(\gamma)+S_{\stat}(\sigma)+\Tr[\nabla S_{\stat}(\sigma)(\gamma-\sigma)],
\end{equation}
the Bregman divergence generated by $-S_{\stat}$. Combining $D(\omega_{\gamma}\|\omega_{\sigma})=D_{\stat}(\gamma\|\sigma)$ with~\eqref{eq:C0} proves the Pythagorean identity~\eqref{eq:pythagoras}. Taking $\omega=\omega_{\gamma}$ in~\eqref{eq:pythagoras} gives $D(\omega_{\gamma}\|\omega_{\sigma})=D_{\stat}(\gamma\|\sigma)$, and, since $D(\omega\|\omega_{\gamma})\ge0$ with equality if and only if $\omega=\omega_{\gamma}$, it also gives $D_{\stat}(\gamma\|\sigma)=\min\{D(\omega\|\omega_{\sigma}):\gamma_{\omega}=\gamma\}$. $\square$

\textit{Proof of Proposition~\ref{prop:dpi}.}---By hypothesis, $\mathcal{N}(\omega_{\gamma})=\omega_{A\gamma A^{\dagger}+B}$ and $\mathcal{N}(\omega_{\sigma})=\omega_{A\sigma A^{\dagger}+B}$. Hence Proposition~\ref{prop:relent}, applied to the outputs and to the inputs, and the data-processing inequality for the Umegaki relative entropy~\cite{Lindblad1975} give
\begin{equation}
D_{\stat}\big(A\gamma A^{\dagger}+B\,\big\|\,A\sigma A^{\dagger}+B\big)=D\big(\mathcal{N}(\omega_{\gamma})\,\big\|\,\mathcal{N}(\omega_{\sigma})\big)\ \le\ D(\omega_{\gamma}\|\omega_{\sigma})=D_{\stat}(\gamma\|\sigma).
\end{equation}
If the outputs are not interior points of $\mathcal{C}_{\stat}$, note that $B\succeq0$, and $AA^{\dagger}+B\preceq I$ for fermions, as follows by letting $\gamma\to0$ (and $\gamma\to I$) in $A\gamma A^{\dagger}+B\in\mathcal{C}_{\stat}$. Since $\gamma$ and $\sigma$ are interior points, both outputs then have the same kernel, $\ker A^{\dagger}\cap\ker B$, and, for fermions, the same eigenvalue-one eigenspace, $\ker A^{\dagger}\cap\ker(I-B)$. The modes in these subspaces are empty (or filled) in both output states, as in the definition of $\omega_{\gamma}$ for boundary $\gamma$ in \appref{A}, and contribute zero to both sides. $\square$

\textit{Channels on $\Hil$.}---We prove the statements made after Proposition~\ref{prop:dpi}.

(a) \emph{Data processing for unnormalized operators.} Let $P,R\succeq0$ with $\operatorname{supp}P\subseteq\operatorname{supp}R$, let $p\coloneqq\Tr[P]>0$ and $r\coloneqq\Tr[R]$, and write $P=p\,\rho$ and $R=r\,\tau$ with $\rho,\tau$ density operators. Then
\begin{equation}
D(P\|R)=p\,\Tr\big[\rho\big(\ln(p\rho)-\ln(r\tau)\big)\big]=p\,D(\rho\|\tau)+p\ln(p/r).
\end{equation}
A channel $\Phi$ is linear and trace preserving, so that $\Phi(P)=p\,\Phi(\rho)$ and $\Phi(R)=r\,\Phi(\tau)$, and the data-processing inequality for density operators~\cite{Lindblad1975} gives
\begin{equation}\label{eq:C-dpi}
D\big(\Phi(P)\,\big\|\,\Phi(R)\big)=p\,D\big(\Phi(\rho)\,\big\|\,\Phi(\tau)\big)+p\ln(p/r)\ \le\ p\,D(\rho\|\tau)+p\ln(p/r)=D(P\|R).
\end{equation}

(b) \emph{Joint convexity.} Let $(p_{i})_{i}$ be a probability vector and $P_{i},R_{i}\succeq0$. The block-diagonal operators $\hat P\coloneqq\sum_{i}p_{i}|i\rangle\!\langle i|\otimes P_{i}$ and $\hat R\coloneqq\sum_{i}p_{i}|i\rangle\!\langle i|\otimes R_{i}$ satisfy
\begin{equation}
D(\hat P\|\hat R)=\sum_{i}D(p_{i}P_{i}\|p_{i}R_{i})=\sum_{i}p_{i}\,D(P_{i}\|R_{i}),
\end{equation}
and the partial trace over the register $\{|i\rangle\}$, which maps them to $\sum_{i}p_{i}P_{i}$ and $\sum_{i}p_{i}R_{i}$, is a channel. Hence~\eqref{eq:C-dpi} gives
\begin{equation}\label{eq:C-jc}
D\Big(\sum_{i}p_{i}P_{i}\,\Big\|\,\sum_{i}p_{i}R_{i}\Big)\ \le\ \sum_{i}p_{i}\,D(P_{i}\|R_{i}).
\end{equation}

(c) \emph{Fermi--Dirac divergence.} If $\Phi$ is a unital channel, $\Phi(I)=I$, then $0\preceq\gamma\preceq I$ implies $0\preceq\Phi(\gamma)\preceq\Phi(I)=I$, and $\Phi(I-\gamma)=I-\Phi(\gamma)$. Applying~\eqref{eq:C-dpi} to each of the two terms of $D_{\mathrm{FD}}$,
\begin{align}
D_{\mathrm{FD}}\big(\Phi(\gamma)\,\big\|\,\Phi(\sigma)\big)&=D\big(\Phi(\gamma)\,\big\|\,\Phi(\sigma)\big)+D\big(\Phi(I-\gamma)\,\big\|\,\Phi(I-\sigma)\big)\notag\\
&\le D(\gamma\|\sigma)+D(I-\gamma\|I-\sigma)=D_{\mathrm{FD}}(\gamma\|\sigma).
\end{align}
Likewise, since $I-\sum_{i}p_{i}\gamma_{i}=\sum_{i}p_{i}(I-\gamma_{i})$, applying~\eqref{eq:C-jc} to each of the two terms gives
\begin{equation}
D_{\mathrm{FD}}\Big(\sum_{i}p_{i}\gamma_{i}\,\Big\|\,\sum_{i}p_{i}\sigma_{i}\Big)\ \le\ \sum_{i}p_{i}\,D_{\mathrm{FD}}(\gamma_{i}\|\sigma_{i}).
\end{equation}

(d) \emph{Bose--Einstein divergence.} In $D_{\mathrm{BE}}(\gamma\|\sigma)=D(\gamma\|\sigma)-D(\gamma+I\|\sigma+I)$, the second term enters with a minus sign, so that~\eqref{eq:C-dpi} and~\eqref{eq:C-jc} imply nothing, and data processing can indeed fail. Consider the mixed-unitary channel $\Phi(Z)=\frac{1}{2}(Z+VZV^{\dagger})$ on $\mathbb{C}^{2}$, with $V=|0\rangle\!\langle1|+|1\rangle\!\langle0|$, and the operators $X=\operatorname{diag}(0.05,50)$ and $Y=\operatorname{diag}(10^{-3},5)$, for which $\Phi(X)=25.025\,I$ and $\Phi(Y)=2.5005\,I$. For commuting arguments, $D_{\mathrm{BE}}$ is the sum over eigenvalues of $d_{\mathrm{BE}}(x\|y)\coloneqq x\ln(x/y)-(x+1)\ln[(x+1)/(y+1)]$, so that
\begin{align}
D_{\mathrm{BE}}(X\|Y)&=d_{\mathrm{BE}}(0.05\|10^{-3})+d_{\mathrm{BE}}(50\|5)\approx0.145+5.986\approx6.13,\\
D_{\mathrm{BE}}\big(\Phi(X)\,\big\|\,\Phi(Y)\big)&=2\,d_{\mathrm{BE}}(25.025\|2.5005)\approx10.86\ >\ D_{\mathrm{BE}}(X\|Y). \qquad\square
\end{align}

\end{document}